\documentclass[twocolumn]{aastex631}
\usepackage{xspace}
\usepackage{subfigure}

\newcommand{\Gaia}{\textit{Gaia}\xspace}
\newcommand{\TESS}{\textit{TESS}\xspace}
\newcommand{\Kepler}{\textit{Kepler}\xspace}

\usepackage[normalem]{ulem}
\newcommand\out{\bgroup\markoverwith
{\textcolor{red}{\rule[.5ex]{2pt}{1pt}}}\ULon}
\usepackage{xcolor}

\shorttitle{Searching for Compact Objects from LAMOST Survey}
\shortauthors{Qi et al.}

\begin{document}

\title{Searching for Compact Object Candidates from LAMOST Time-Domain Survey of Four K2 Plates}

\correspondingauthor{Wei-Min Gu, Tuan Yi, Zhi-Xiang Zhang}
\email{guwm@xmu.edu.cn \\
yit@xmu.edu.cn \\
zhangzx@xmu.edu.cn}

\author[0000-0002-7135-6632]{Senyu Qi}
\affiliation{Department of Astronomy, Xiamen University, Xiamen, Fujian 361005, China}

\author[0000-0003-3137-1851]{Wei-Min Gu}
\affiliation{Department of Astronomy, Xiamen University, Xiamen, Fujian 361005, China}

\author[0000-0002-5839-6744]{Tuan Yi}
\affiliation{Department of Astronomy, Xiamen University, Xiamen, Fujian 361005, China}

\author[0000-0002-2419-6875]{Zhi-Xiang Zhang}
\affiliation{Department of Astronomy, Xiamen University, Xiamen, Fujian 361005, China}

\author[0000-0003-3116-5038]{Song Wang}
\affiliation{National Astronomical Observatories, Chinese Academy of Sciences, Beijing 100101, China}

\author[0000-0002-2874-2706]{Jifeng Liu}
\affiliation{National Astronomical Observatories, Chinese Academy of Sciences, Beijing 100101, China}
\affiliation{College of Astronomy and Space Science, University of Chinese Academy of Sciences, Beijing 100101, China}

\begin{abstract}
 
The time-domain (TD) surveys of the Large Sky Area Multi-Object Fiber Spectroscopic Telescope (LAMOST) yield high-cadence radial velocities, paving a new avenue to study binary systems including compact objects.
In this work, we explore LAMOST TD spectroscopic data of four K2 plates and present a sample of six single-lined spectroscopic binaries that may contain compact objects.
We conduct analyses using phase-resolved radial velocity measurements of the visible star, to characterize each source and to infer the properties of invisible companion. By fitting the radial velocity curves for the six targets, we obtain accurate orbital periods, ranging from $\sim$ (0.6--6) days, and radial velocity semi-amplitudes, ranging from $\sim$ (50--130) km s$^{-1}$. 
We calculate the mass function of the unseen companions to be between 0.08 and 0.17 $M_{\odot}$. Based on the mass function and the estimated stellar parameters of the visible star, we determine the minimum mass of the hidden star.
Three targets, J034813, J063350, and J064850, show ellipsoidal variability in the light curves from K2, ZTF, and \textit{TESS} surveys.
Therefore, we can put constraints on the mass of the invisible star using the ellipsoidal variability. 
We identify no X-ray counterparts for these targets except for J085120, of which the X-ray emission can be ascribed to stellar activity.
We note that the nature of these six candidates is worth further characterization utilizing multi-wavelength follow-up observations.

\end{abstract}


\section{Introduction} \label{sec:intro}

White dwarfs (WDs), neutron stars (NSs), and black holes (BHs) are remnants of stars that ended their lifetimes. 
The identification and characterization of these objects can yield key insights into strong gravitational fields, 
stellar formation, evolution history, binary interactions and evolution, and the properties and behavior of matter at extreme densities. 
Additionally, compact objects are closely related to various astronomical phenomena, such as novae, kilonovae, supernovae, X-ray bursts, gamma-ray bursts \citep{2022Natur.612..228T}, and gravitational waves.
Many observations show that more than half of stars in the Galaxy are in binary systems
\citep{2013ARA&A..51..269D} and that binary systems contain a substantial number of compact objects
\citep{2006ARA&A..44...49R,1984ARA&A..22..537J,2012MNRAS.419..806R}.

The number of binaries including WDs has skyrocketed in recent years, thanks to state-of-the-art time-domain surveys
\citep{2016MNRAS.458.3808R,2016MNRAS.463.2125P,2018MNRAS.477.4641R,2021MNRAS.505.2051E,2021MNRAS.506.5201R,2022ApJ...933..193Z,2022ApJ...936...33Z}. The identification of the majority of the white dwarf–main-sequence (WD-MS) binaries relies on the distinctive spectral characteristics of WDs. 
And the majority of binary systems containing NSs or stellar-mass BHs are discovered by their bright bursts in the X-ray band
\citep{2012ApJ...757...36K,2014Natur.505..378C,2016A&A...587A..61C}.
The X-ray outbursts are caused by mass transfer from the companion overflowing the Roche lobe or stellar wind coming from the companion.
Recent gravitational wave detection has confirmed some compact object binaries
\citep{2016PhRvL.116f1102A,2017ApJ...848L..12A,2020ApJ...892L...3A}.
However, these sources only account for a small population of total compact object binaries. Stellar evolution theory predicts that there are about $\sim 10^8$ stellar-mass BHs and $\sim 10^9$ NSs in our Milky Way
\citep{1994ApJ...423..659B}. This indicates that there are still plenty of undiscovered quiescent dark objects in our galaxy. 

One promising strategy for discovering more dormant compact objects is to search for single-lined spectroscopic binary systems with large mass functions. The methodology, commonly known as the dynamical method, utilizes multi-epoch spectroscopic observations to monitor stars with large radial velocity variations potentially induced by a compact companion. Dynamical searches for non-interacting BHs binaries and NSs binaries \citep{1969ApJ...156.1013T} commenced even before the identification of the first BH in X-ray binaries.
Recently, a couple of stellar-mass BHs and NSs in binary systems have been discovered based on the dynamical method
\citep{2019Sci...366..637T,2019Natur.575..618L,2020ApJ...900...42L,2021MNRAS.504.2577J,2022MNRAS.tmp.2933E,2022NatAs.tmp..201Y,2022arXiv221004685Z,2023arXiv230207880E}. 
Using time-domain (TD) spectroscopic, astrometric, and photometric surveys, the orbital parameters of the systems and the stellar properties of the optically visible stars can be determined, and the presence of hidden companions can be dynamically confirmed. 

The Large Sky Area Multi-Object Fiber Spectroscopic Telescope (LAMOST) is a reflecting Schmidt telescope with a four-meter effective aperture and a wide field of view \citep{2012RAA....12.1197C}. 
LAMOST provides millions of stellar spectra in both medium-resolution mode ($R \sim 7500$) and low-resolution mode ($R\sim 1800$).
LAMOST's observation strategy of taking multiple spectra for individual targets provides an ideal opportunity to search for binary systems containing compact objects by dynamical methods.
Furthermore, numerous ground- or space-based photometric surveys are now providing vast quantities of photometric data, which include the \Kepler K2 mission \citep{2014PASP..126..398H}, the Zwicky Transient Facility
\cite[ZTF;][]{2019PASP..131a8002B}, the All-Sky Automated Survey for Supernovae 
\cite[ASAS-SN;][]{2014ApJ...788...48S} and Transiting Exoplanet Survey Satellite 
\cite[\TESS;][]{2015JATIS...1a4003R}. The high-cadence photometry has the potential to be an effective means for searching for compact object candidates
\citep{2021MNRAS.507..104R,2021MNRAS.501.2822G,2022arXiv220606032G}.
Utilizing the TD spectroscopic data from LAMOST and the wealth of photometric data from ground- or space-based surveys, a dozen of compact object candidates have been revealed \citep{2019ApJ...872L..20G,2019AJ....158..179Z,2021ApJ...923..226Y,2022SCPMA..6529711M,2022ApJ...938...78L,2022MNRAS.517.4005M,2022ApJ...940..165Y}.

This paper aims to search for compact objects in binary systems based on the LAMOST TD spectroscopic data of four K2 plates \citep[][see their Figure 1]{2021RAA....21..292W}.
We select a sample containing six single-lined spectroscopic binaries which all have more than 20 radial velocity measurements. We determine the orbital parameters from radial velocity curves and light curves. Then the mass of the optically invisible stars in binaries is estimated by using the mass function. We introduce the details of the data and the criterion for selecting candidates in Section \ref{sec:data}. The properties of the sample are shown in Section \ref{sec:result}, including the orbital parameters and the stellar parameters. The discussion and the summary are presented in Section \ref{sec: the discussion} and Section \ref{sec: summmary}, respectively.

\section{Data selection} 

\label{sec:data}

\subsection{Data} 
From Oct. 2019 to Apr. 2020, LAMOST conducted a TD survey to observe four footprints (fields) of the K2 campaign \cite[][we refer to the survey as four K2 plates]{2021RAA....21..292W}. 
The survey extends the achievements of the LAMOST-Kepler/K2 projects conducted TD and non-TD sky surveys covering the Kepler field and the K2 campaigns from 2012 to 2019 \citep{2020RAA....20..167F,2020ApJS..251...15Z}.
The four K2 plates have collected $\sim$767,000 low-resolution and $\sim$478,000 median-resolution spectra.
\citet{2021RAA....21..292W} used different methods, including the LAMOST Stellar Parameter Pipeline (LASP), the DD-Payne, and the Stellar LAbel Machine (SLAM) to determine the stellar parameters, namely, effective temperature, surface gravity, metallicity, and radial velocity. We select 7245 sources from the sample with multiple med-resolution observations and a mean signal-to-noise ratio (SNR) greater than 10. In our sub-sample, 5468 sources have more than 10 med-resolution exposures and 4121 sources have more than 20 exposures. 
All these data are available in the LAMOST DR9\footnote{\url{http://www.lamost.org/dr9/v1.0/catalogue}}.

\begin{deluxetable*}{lrrll}
\tablecaption{Basic information for the sources in our sample.\label{table:1}}
\tablewidth{0.99\textwidth}
\tabletypesize{\scriptsize}
\tablehead
{
\colhead{ID} & \colhead{R.A.} & \colhead{Decl} & \colhead{$\varpi$} & \colhead{gmag} \\
\colhead{} & \colhead{(J2000)} & \colhead{(J2000)} & \colhead{(mas)} & \colhead{(mag)}
}
\startdata
J034813 & 57.055792 & 25.098805 & $1.168\pm0.013$ & 12.24\\
J063350 & 98.459806 & 22.015557 & $0.304\pm0.026$ & 14.54\\
J064717 & 101.822211 & 24.609382 & $0.399\pm0.020$ & 13.96\\
J064850 & 102.209808 & 24.830386 & $0.711\pm0.018$ & 13.89\\
J085102 & 132.758841 & 11.817029 & $1.280\pm0.048$ & 13.17\\
J104734 & 161.894238 & 7.923628  & $1.729\pm0.015$ & 13.15\\
\enddata
\tablecomments{Column (1): designation of the source; 
column (2): R.A. (J2000); 
column (3): Decl.(J2000); 
column (4): parallax from \Gaia\ DR3; 
column (5): g-band magnitude from \Gaia\ DR3.}
\end{deluxetable*}

\subsection{Candidates Selection}

Compact object candidates can be selected using the mass function:
\begin{equation}\label{eq1}
f(M_2) \equiv \frac{M_2^{3}\sin^{3}i}{{\left(M_1+M_2\right)}^2} = \frac{K^3_1P_{\rm{orb}}}{2\pi G},
\end{equation}
where $M_1$ is the mass of the optically visible star and $M_2$ is the mass of the invisible companion, $K_1$ is the radial velocity semi-amplitude of the visible star, $P_{\rm{orb}}$ is the orbital period, $i$ is the orbital inclination angle, and \textit{G} is the gravitational constant. 
The mass function is a stringent lower mass limit for the unseen companion when $i = 90^{\circ}$ and $M_{1} = 0 M_{\odot}$.
$K_1$ and $P_{\rm{orb}}$ can be measured by fitting the radial velocity curve and the light curve.
To find promising compact object candidates, we want sources with large mass functions. It is then obvious from Equation~(\ref{eq1}) that sources with either (or both) large $K_{1}$ or $P_{\rm orb}$ are favored.

According to previous research on binary systems containing compact objects, such as some stellar-mass BH binaries, a high proportion of the companion stars in these systems exhibit large radial velocity changes \citep{2016A&A...587A..61C}. And the distribution of their orbital periods ranges from a few hours to tens of days \citep{2013ApJ...768..185S,2013Sci...339.1048C}. 

In the first phase, we pick sources that have ${\Delta}V_{\rm{R}}\  {\gtrsim}\  80 \ \mathrm{km \ s^{-1}}$ and SNR ${\geq}\ 10$, where ${\Delta}V_{\rm{R}}$ is the largest radial velocity variation among all spectroscopic data for a given source. 
Then we use the Lomb-Scargle algorithm \citep{1976Ap&SS..39..447L,1981ApJS...45....1S} to search for periodic signals from the radial velocity measurements. The period corresponding to the highest Lomb-Scargle power is used to phase-fold the radial velocity curve.
We visually inspect each radial velocity curve, discard poorly folded curves (e.g., sources with no significant periodic variation), and retain a sample of well-folded curves.

Note that each of our sources has got covered by at least 20, up to 60 repeating spectroscopic observations, ensuring that the radial velocity measurements encompass most of the orbital phases. Therefore, ${\Delta}V_{\rm{R}}/2$ is a good approximation of $K_1$ and is used to estimate the mass function. Based on the ${\Delta}V_{\rm{R}}/2$ and the period, we evaluate the mass functions by using Equation~(\ref{eq1}) and select the sources with $f(M_2)\gtrsim 0.1 M_{\odot}$.

After that, we check if the spectrum of each source is single-lined. If a source has a single-lined spectrum but an apparent radial velocity variation, an unseen star that is either a compact object or a much fainter star may exist in this binary system.
We employ the method 
\citep{2017A&A...608A..95M,2021ApJS..256...31L} based on detecting multiple peaks of the cross correction function (CCF) successive derivatives to determine double-lined (and multi-lined) spectroscopic systems (see Appendix). These non-single-lined spectroscopic binaries are removed from our sample. 

Before reaching a final sample of candidates, we cross-match these sources with the K2 mission \citep{2014PASP..126..398H}, ZTF \citep{2019PASP..131a8002B}, ASAS-SN \citep{2014ApJ...788...48S}, and \textit{TESS} \citep{2015JATIS...1a4003R}. By using the light curves obtained from these surveys, we can exclude false positives such as close eclipsing binaries which obviously consist of two non-compact stars.

Finally, we gain a sample of six\footnote{When assembling our sample, we found three compact object candidates have been proposed by \citet{2022ApJ...938...78L} using the same LAMOST-K2 TD survey, therefore we do not include these three targets in our sample.}
single-lined spectroscopic binary with significant radial velocity variations as candidates who may host compact objects. Their basic information are presented in Table~\ref{table:1}.

\begin{deluxetable*}{lllrlllrrlll}
\tablecaption{Stellar parameters for the sources in our sample.\label{table:2}}
\tablewidth{0.99\textwidth}
\tabletypesize{\scriptsize}
\tablehead
{
\colhead{ID} & \colhead{$P_{\rm{orb}}$} & \colhead{$P_{\rm{ph}}$} & \colhead{$K_{1}$}  & \colhead{$e$}  & \colhead{$R_1$} & \colhead{$T_{\rm eff}$} & \colhead{log$g$}  & \colhead{$M_{1}$}  & \colhead{$f$} & \colhead{$f(M_{2})$} & \colhead{$M^{\rm{min}}_{2}$}  \\
\colhead{} & \colhead{(day)} & \colhead{(day)} & \colhead{($\mathrm{km\,s^{-1}}$)}  & \colhead{} & \colhead{($R_{\odot}$)} & \colhead{(K)} & \colhead{(dex)}  & \colhead{($M_{\odot}$)} & \colhead{($ \equiv R_1/R_{\rm{L1}}$)} & \colhead{($M_{\odot}$)} & \colhead{($M_{\odot}$)} 
}
\startdata
J034813 & 3.93 & $3.9289$ & $61.72\pm0.30$ & $0.002^{+0.002}_{-0.001}$ &$3.19^{+0.16}_{-0.13}$ & $5440^{+65}_{-60}$ & $3.62^{+0.10}_{-0.09}$ & $1.35^{+0.21}_{-0.16}$ & $0.60^{+0.03}_{-0.04}$ & $0.096\pm0.001$ & $0.75^{+0.07}_{-0.06}$ \\
J063350 & 0.69 & $0.6934$ & $133.22\pm0.47$ & $0.069^{+0.006}_{-0.008}$ & $2.36^{+0.09}_{-0.08}$ & $7644^{+80}_{-67}$ & $4.15^{+0.07}_{-0.07}$ & $^{*}1.74^{+0.20}_{-0.20}$ & $^{*}1.30^{+0.07}_{-0.07}$ & $0.170\pm0.002$ & $1.11^{+0.07}_{-0.08}$\\
J064717 & 2.91 & $--$ & $67.18\pm0.40$ & $0.012^{+0.005}_{-0.005}$ & $2.10^{+0.33}_{-0.17}$ & $6802^{+33}_{-33}$ & $3.93^{+0.07}_{-0.05}$ & $1.43^{+0.23}_{-0.18}$ & $0.47^{+0.04}_{-0.08}$ & $0.092\pm0.002$ & $0.76^{+0.07}_{-0.06}$\\
J064850 & 1.23 & $1.2292$ & $99.10\pm0.64$ & $0.041^{+0.006}_{-0.005}$ & $1.99^{+0.12}_{-0.11}$ & $5906^{+55}_{-48}$ & $3.96^{+0.09}_{-0.12}$ & $1.09^{+0.12}_{-0.11}$ & $0.87^{+0.06}_{-0.06}$ & $0.124\pm0.002$ & $0.75^{+0.05}_{-0.05}$\\
J085102 & 5.78 & $--$ & $53.57\pm0.29$ & $0.003^{+0.003}_{-0.002}$ & $1.46^{+0.10}_{-0.10}$ & $5880^{+23}_{-27}$ & $3.95^{+0.13}_{-0.06}$ & $1.05^{+0.10}_{-0.06}$ & $0.23^{+0.02}_{-0.02}$ & $0.092\pm0.001$ & $0.64^{+0.04}_{-0.03}$\\
J104734 & 4.10 & $--$ & $58.34\pm0.31$ & $0.022^{+0.005}_{-0.005}$ & $1.24^{+0.04}_{-0.04}$ & $5685^{+23}_{-40}$ & $4.24^{+0.06}_{-0.08}$ & $0.91^{+0.05}_{-0.03}$ & $0.26^{+0.01}_{-0.01}$ & $0.084\pm0.001$ & $0.57^{+0.02}_{-0.01}$\\
\enddata
\tablecomments{Column (1): designation of the source; 
column (2): orbital period from radial velocity curve; 
column (3): photometric period from light curve; 
column (4): semi-amplitude of radial velocity curve;
column (5): eccentricity;
column (6): radius measured by SED fitting; 
column (7): effective temperature measured by SED fitting; 
column (8): surface gravity measured by SED fitting; 
column (9): mass from the MIST model; 
column (10): filling factor. The $^{*}$ symbol denotes that the result $f > 1$ is non-physical (see discussion in Section~\ref{subsec:filling}); 
column (11): mass function;
column (12): mass of the second star for $i = 90^{\circ}$.}
\end{deluxetable*}

\section{results}\label{sec:result}

\subsection{Radial velocity curves and mass functions}

We measure the orbital periods from radial velocity measurements using two methods. As mentioned, first, we use the Lomb-Scargle method to search for the periodic signals from the radial velocity measurements.
Second, we use \texttt{TheJoker} to fit the radial velocity curve. \texttt{TheJoker} is a python-based package that implements the Monte Carlo sampling technique to fit the two-body problem \citep{2017ApJ...837...20P}.
A uniform period prior spanning [0.1 d, 20 d] is set to for \texttt{TheJoker}. 
Consistent orbital periods are obtained from these two methods.

\begin{figure*}[t]
\centering
\includegraphics[scale=0.2]{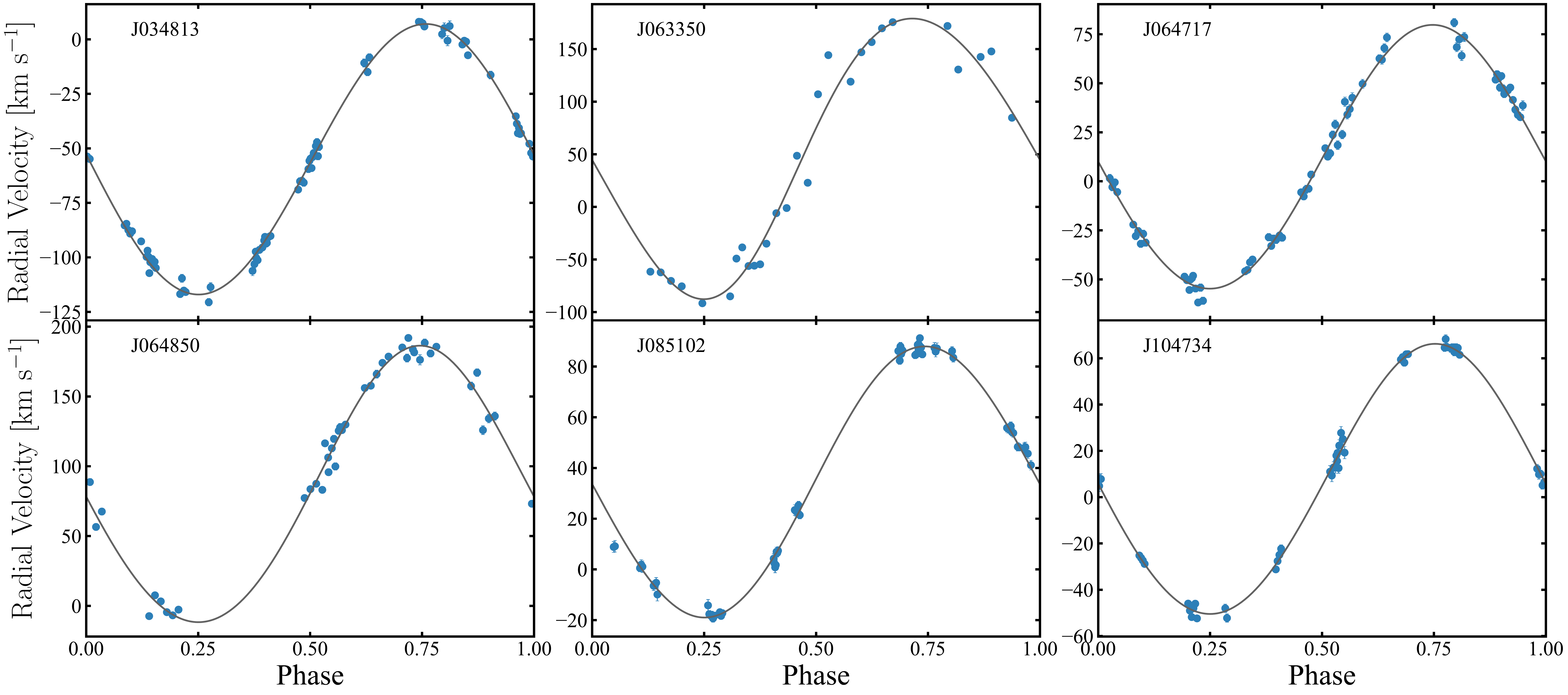}
\caption{Phase-folded radial velocity data (dots) and the best-fitted radial velocity curves (lines) of the selected sources. The error bars for most data points are too small to be seen.}
\label{fig:fig1}
\end{figure*}

Following this, we derive the orbital parameters by fitting the radial velocity curve.
We use the general form of a Keplerian orbit, that is:
$V_{\mathrm{r}}(t) = v_0 + K_1(\cos(f+w)+e\cos(w))$, 
where $v_0$ is the barycentric velocity (center-of-mass radial velocity), $f$ is the true anomaly, $w$ is the argument of periastron, and $e$ is the eccentricity. 
The fitted radial velocity curves are shown in Figure~\ref{fig:fig1} and the radial velocity semi-amplitudes are presented in Table~\ref{table:2}.
Then the mass function of each source is then calculated using Equation \ref{eq1} and the results are also listed in Table~\ref{table:2}.
The fitting results reveal that the eccentricity of all sources is approximately equal to zero, indicating that the orbit is circularized.

\subsection{Light curves}

Meanwhile, we collect the light curves from various photometric TD surveys: K2, \TESS, ZTF\footnote{\url{https://irsa.ipac.caltech.edu/cgi-bin/Gator/nph-scan?projshort=ZTF&mission=irsa}}, and ASAS-SN\footnote{\url{https://asas-sn.osu.edu/variables}}. 
Light curves from K2 and \TESS are obtained using the python package \texttt{Lightkurve} \citep{2018ascl.soft12013L}.
We measure the photometric periods ($P_{\rm{ph}}$) using the Lomb-Scargle algorithm.
Three selected sources have a $P_{\rm{ph}}$ consistent with the orbital periods from the radial velocity curve fitting, and the phased-folded light curves are reliable based on our visual inspections (Figure~\ref{fig:fig3}).
Their light curves show typical ellipsoidal modulation (a quasi-sinusoidal variation with double peaks and double valleys feature). The remaining three sources have no discernible periodic variations.

\subsection{Mass constraints}
We use the broad-band spectral energy distributing (SED) fitting to constrain the stellar parameters of sources by using a python package \texttt{astroARIADNE}\footnote{\url{https://github.com/jvines/astroARIADNE}}.
\texttt{AstroARIADNE} is designed as the program that uses the Nested Sampling algorithm to automatically fit the SED of target stars using as many as six distinct atmospheric model grids, to derive effective temperature, surface gravity, metallicity, distance, radius, and V-band extinction \citep{2022MNRAS.513.2719V}.

We collect multi-band photometric data including GALEX \citep{2005ApJ...619L...1M,2011Ap&SS.335..161B}, SDSS \citep{2009ApJS..182..543A}, APASS \citep{2014CoSka..43..518H}, Pan-STARRS \citep{2016arXiv161205560C}, $TESS$ \citep{2015JATIS...1a4003R,2019AJ....158..138S}, 2MASS \citep{2006AJ....131.1163S} and ALLWISE \citep{2010AJ....140.1868W} to fit the SED. Then we use the parallax from \Gaia DR3
\citep{2022arXiv220800211G} as the prior of distance. We derive the stellar parameters including $T_{\rm{eff}}$, log$g$, [Fe/H], and radius from the SED fitting, which are summarized in Table~\ref{table:2}. Several SED fitting results are displayed in Appendix.

We use the stellar evolution models to evaluate the mass of visible 
companions by utilizing the python package 
\texttt{isochrones}\footnote{\url{https://isochrones.readthedocs.io/en/latest/}}.
We use the SED best-fit parameters and photometry as inputs to derive 
isochrone interpolated mass with MESA Isochrones \& Stellar Tracks 
isochrones \citep[MIST;][]{2016ApJS..222....8D}. The isochrone mass 
is shown in Table~\ref{table:2}.

\begin{figure*}[t]
\centering
\includegraphics[scale=0.40]{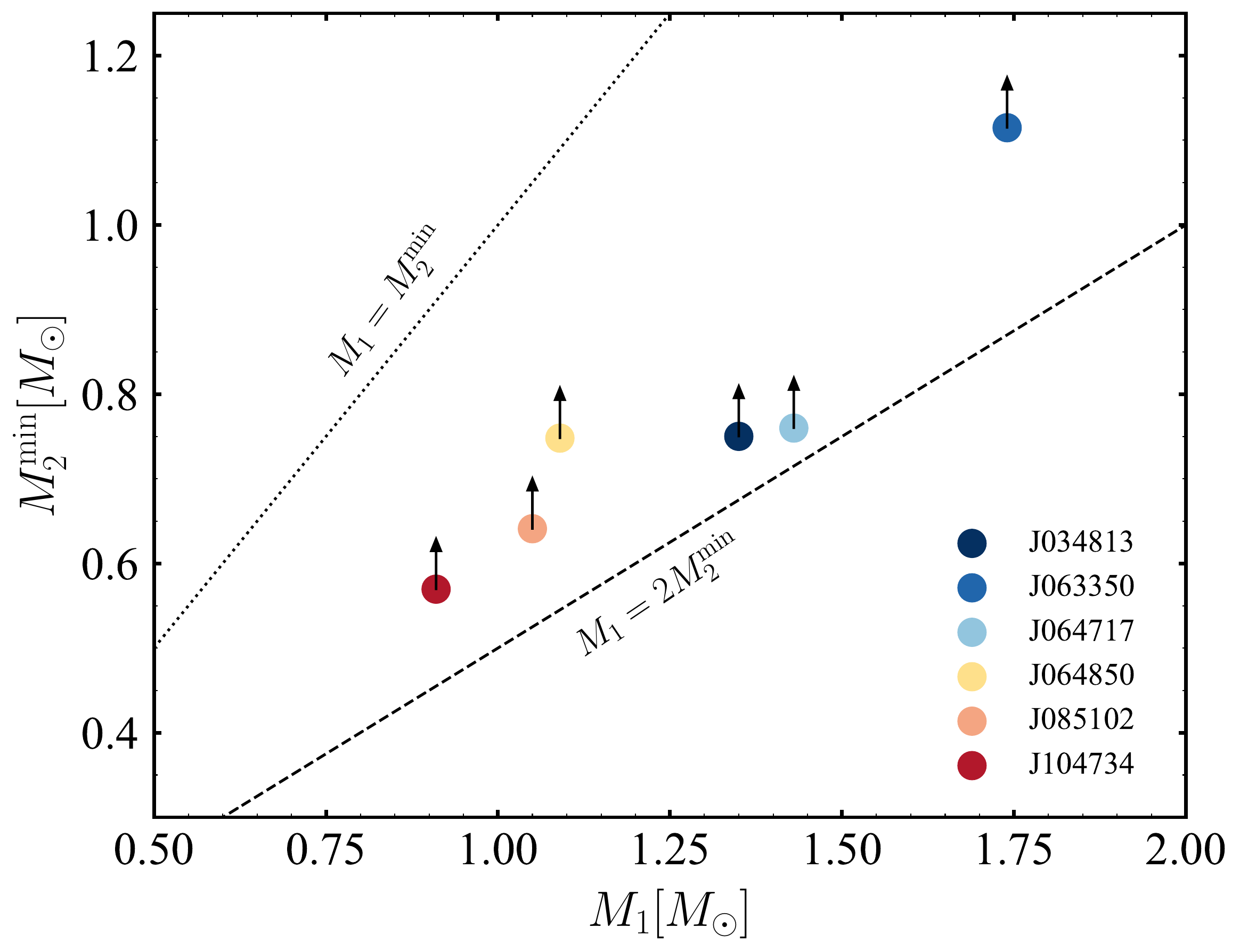}
\caption{The minimum mass of the invisible companion $M^{\rm{min}}_2$ at the maximum inclination ($i=90^{\circ}$). Two dash lines represent the cases $M_1=M^{\rm{min}}_2$ and $M_1=2M^{\rm{min}}_2$.}
\label{fig:fig2}
\end{figure*}

We adopt the isochrone mass as the mass of visible stars to constrain the mass of invisible companions. Combined with the mass function equation, the lower limit of the invisible object's mass $M^{\rm{min}}_2$ is calculated under the assumption that $i=90^{\circ}$. 
We find that $M^{\rm{min}}_2$ exceeds half of the visible stellar mass for all sources. Since a normal stellar companion of similar mass would contribute non-negligible optical flux and manifest double-lined spectroscopic features, the estimation indicates that each of the systems could conceal a compact object.
The result is presented in the right panel of Figure~\ref{fig:fig2}.

\section{discussion}
\label{sec: the discussion}

\subsection{Filling factor}\label{subsec:filling}

Three of our chosen candidates exhibit standard ellipsoidal variability, whereas the remaining sources lack periodic photometric variability. After determining their binary orbital parameters and observable star parameters, we calculate the Roche lobe filling factor to characterize each system. The Roche lobe filling factor is defined as $f \equiv R_1/R_{\rm{L1}}$, where the $R_{\rm{L1}}$ is the Roche lobe radius of the visible star. 
We express the $R_{\rm{L1}}$ as the form of \citep{1971ARA&A...9..183P} 
\begin{equation}\label{eq2}
{\frac{R_{\rm{L1}}}{a}} = 0.462 \left(\frac{M_1}{M_1 \ + \ M_2}\right)^{1/3},
\end{equation}
where $a$ is the binary separation.
The Kepler’s third law shows the relation between $a$ and $P_{\mathrm{orb}}$, which take the form as
\begin{equation}\label{eq3}
    \frac{G(M_1+M_2)}{a^3} = \frac{4{\pi}^2}{P_{\mathrm{orb}}^2}.
\end{equation}
By combining the Equation~(\ref{eq2}) and the Equation~(\ref{eq3}) we obtain
\begin{equation}\label{eq4}
\frac{M_{1}}{R^{3}_{\rm{L1}}} = 0.804 \ P^{-2}_{\rm{day}}\ {\rm g\,cm^{-3}},
\end{equation}
where $P_{\rm{day}}$ is the orbital period in the unit of days.
We use Equation~(\ref{eq4}) to estimate $R_{\rm{L1}}$. The filling factor is then estimated by $f \equiv R_{1} / R_{\rm{L1}}$, where $R_{1}$ is measured from SED fitting. The results are listed in Table~\ref{table:2}.
The result is consistent in that sources with larger $f$ exhibit significant periodic light curves, such as J034813 and J064850. Sources with small $f$ are still far from filling their Roche lobe radii, therefore, they do not exhibit periodic photometric fluctuations.

The filling factor of J066350 is greater than 1, which is a non-physical picture since the Roche lobe radius cannot fall below the physical stellar radius.
The photometric variability of J066350 reaches approximately 20\% in amplitude, 
indicating that the optically visible star is close to or even fills up the Roche lobe. 
We suspect that the system has undergone a mass transfer process, 
and it may be inappropriate to estimate the stellar parameters by using stellar evolution models. 
We use $\log g$ values sampled by \texttt{astroARIADNE} to find that $M^{\rm{grav}}_1=10^{{\rm{log}}g}{R_1}^2 /G  = 2.8^{+0.9}_{-0.7} M_{\odot}$. 
In this case, the corresponding filling factor is around 0.94 to 1.22, suggesting that the mass uncertainty may be the cause of the $f>1$ result 
and the parameters of this object could be physically consistent.

\begin{figure*}[t]
\centering
\includegraphics[scale=0.23]{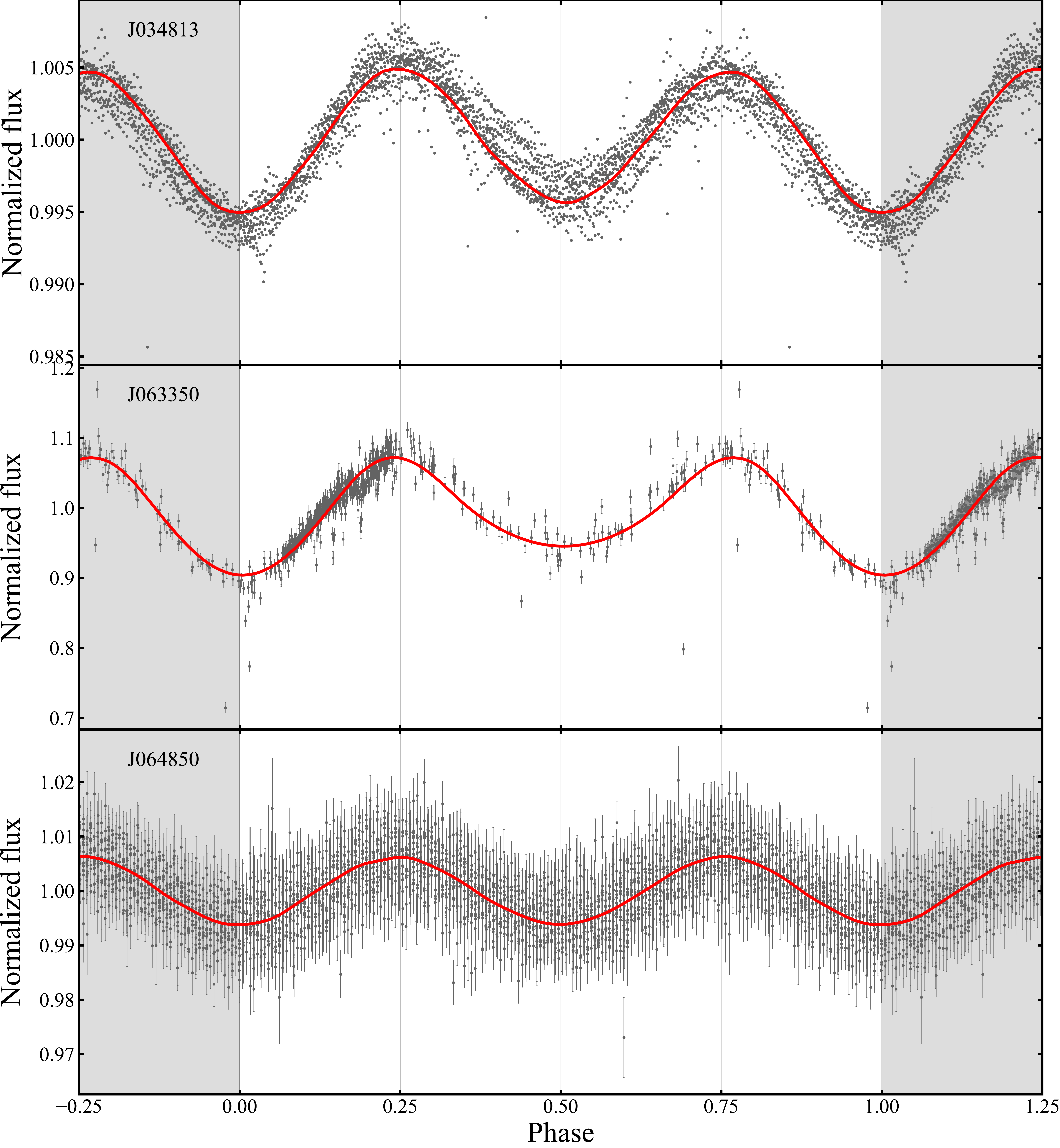}
\caption{Light curves of three sources folded by orbital periods. The photometric data of J034813 is from K2 (The error bars for data points are invisibly small). The photometric data of J063350 is from ZTF. The photometric data of J064850 is from \TESS. The red curve represents the best-fitting \texttt{PHOEBE} model.}
\label{fig:fig3}
\end{figure*}

\subsection{Constraining the invisible object's mass}\label{Sec:4.2}

So far, we have only obtained a lower mass limit for the unseen component using the mass function, i.e., assuming that the orbital inclination $i = 90 ^{\circ}$. In this section, we make further constraints for the candidates showing ellipsoidal light curves.

The three selected candidates, J034813, J063350, and J064850, display typical ellipsoidal variability with a double-peaked profile. The cause of this variability is the tidal interaction between the binary system's objects, which distorts the shape of the stars. 
The key factors that affect the ellipsoidal variability are the binary inclination angle $i$, the mass ratio $q = M_{\rm{2}}/M_{\rm{1}}$, the Roche lobe filling factor $f$, the limb-darkening coefficient, and the gravity-darkening coefficient. 
To determine the values of $i$ and $q$, we utilize the \texttt{PHOEBE 2.4} \citep{2005ApJ...628..426P,2011ascl.soft06002P,2020ApJS..250...34C} software to fit the light curves. The procedures are outlined below.

\begin{figure*}[t]
\centering
\includegraphics[width=0.49\textwidth]{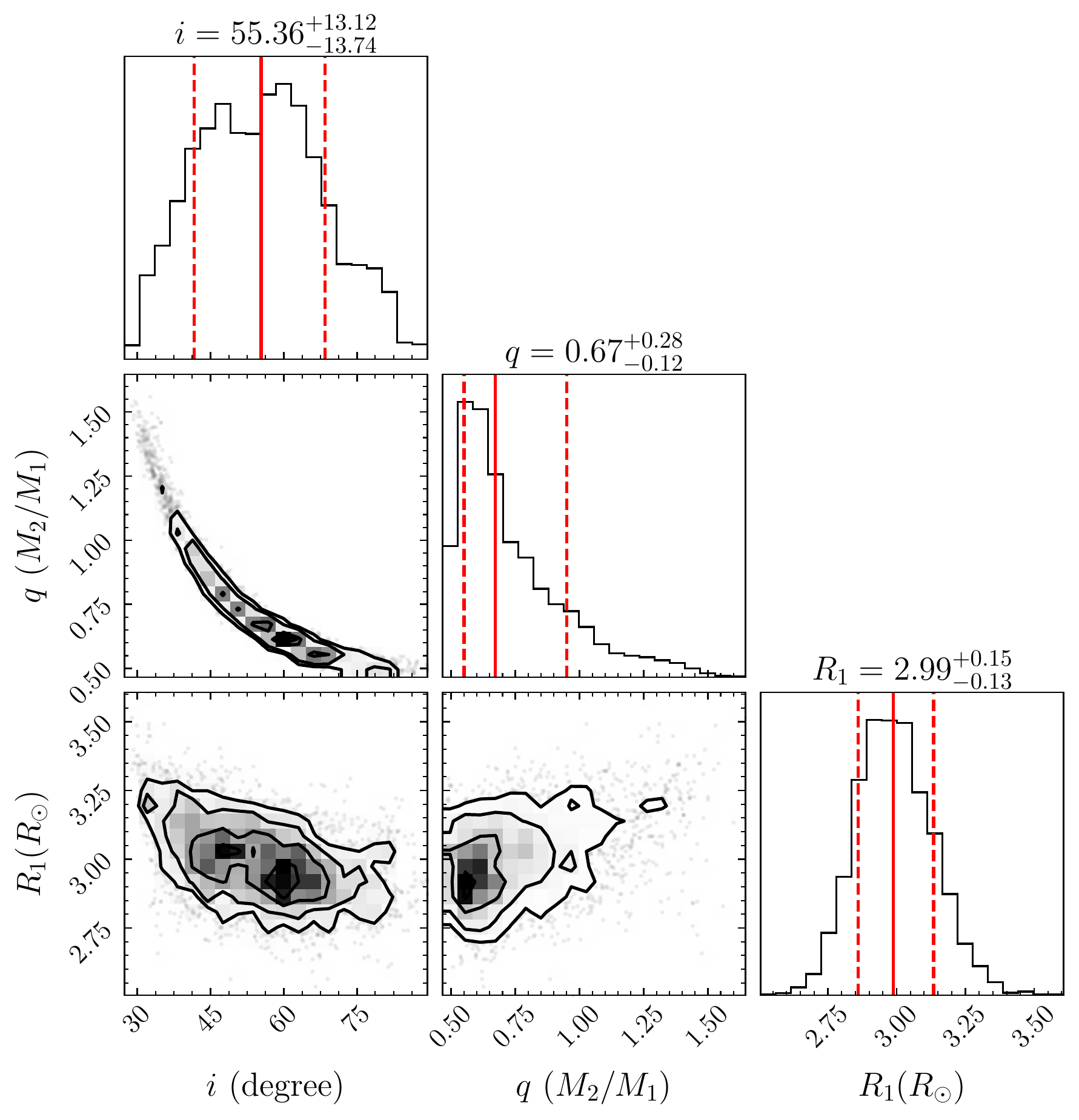}
\includegraphics[width=0.49\textwidth]{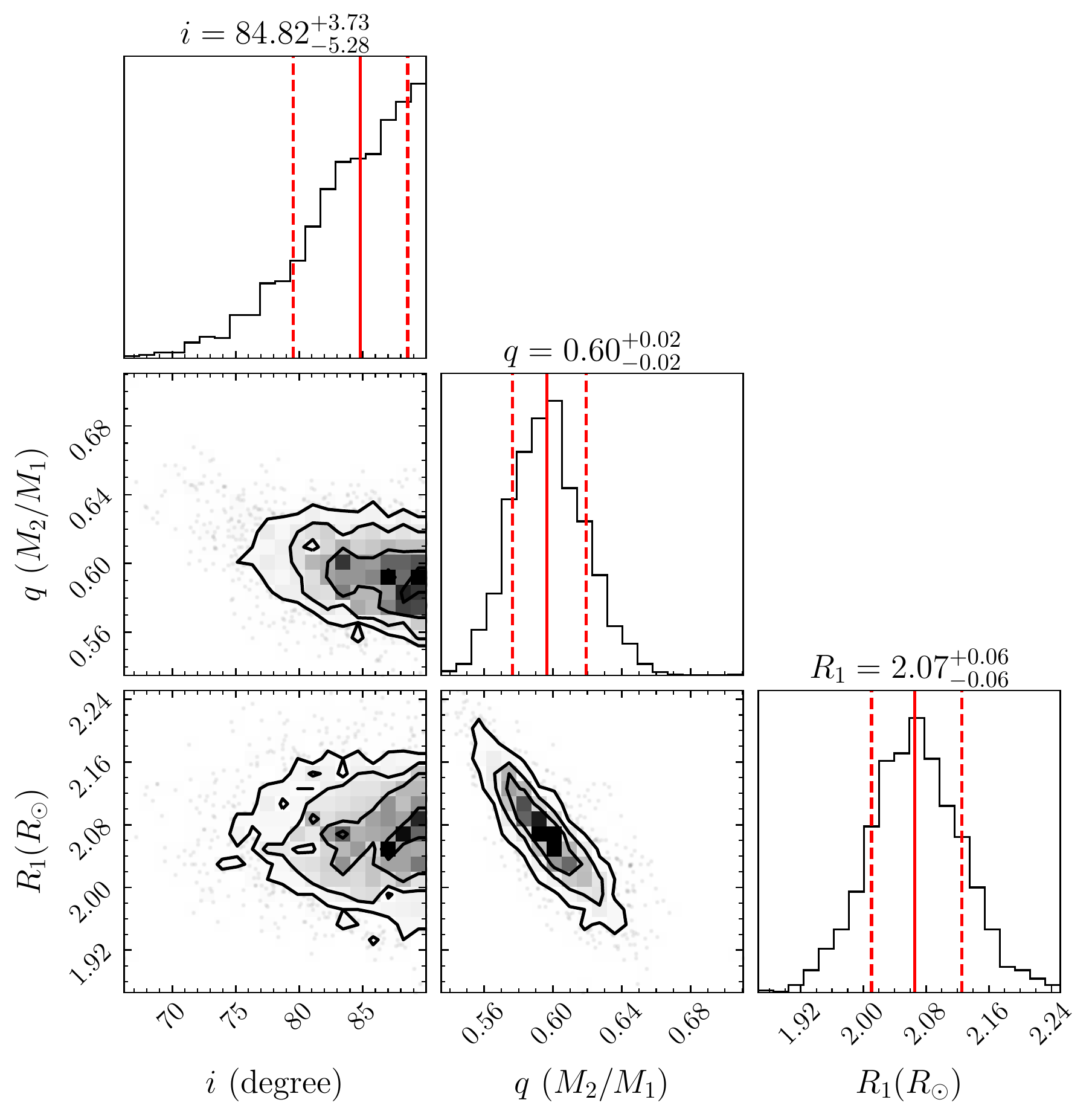}
\includegraphics[width=0.49\textwidth]{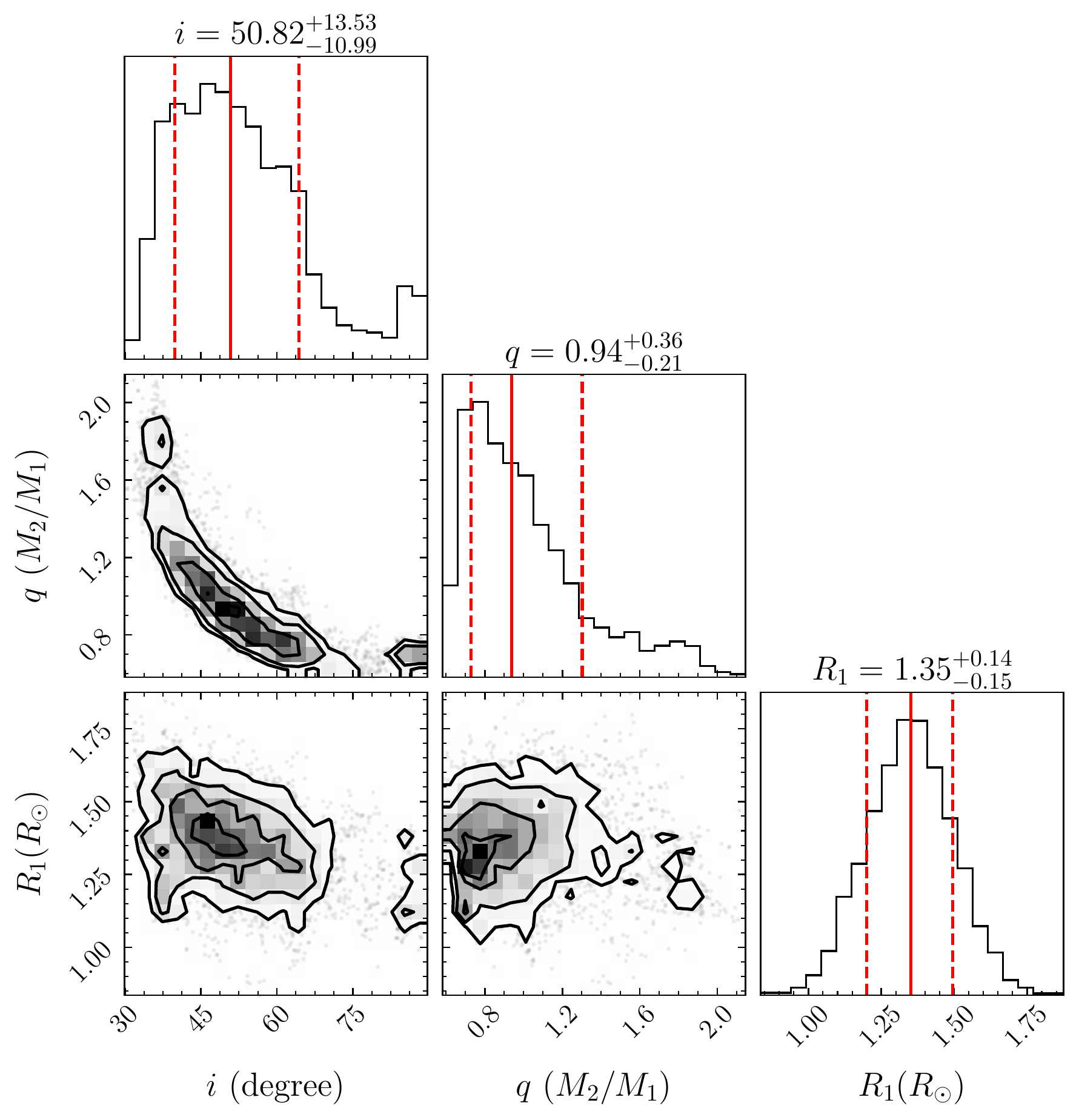}
\caption{Parameter distributions of the light curves fitting. The top left, top right, and bottom corners correspond to J034813, J063350, and J064850, respectively.}
\label{fig:fig4}
\end{figure*}

When fitting the light curves, we keep the orbital period and effective temperature of the optically dominant star fixed at the values listed in Table \ref{table:2}. 
We set \texttt{distortion method $=$ none} to model the invisible companion as an object without any contributions to flux or eclipses. The limb-darkening coefficients are derived from the \texttt{PHOEBE} atmosphere model with a logarithmic limb-darkening law. 
The gravitational darkening coefficient is treated as a free parameter and is assigned a prior of $\beta_1 \sim \mathcal{N}(0.32, 0.1)$ \citep[see][]{2011A&A...529A..75C,badry2021elm}. To perform the MCMC fitting, we use the radius and mass values provided in Table \ref{table:2} as priors. We run multiple parallel chains using the \texttt{emcee} package \citep{2013PASP..125..306F}, with each chain taking 10000 steps.

Figure \ref{fig:fig4} displays the posterior samples for $q$, $i$, and $R_1$, while the best-fitting \texttt{PHOEBE} model results are illustrated by the red lines in Figure \ref{fig:fig3}. A pure ellipsoidal model adequately fits the observed light curve. 
Our fitting results suggest $i=55.36^{+13.12}_{-13.73}$ degree for J034813, with a corresponding invisible object mass of $M_2=0.98^{+0.46}_{-0.20} M_\odot$. The model indicates that the 
inclination of J063350 is nearly edge-on ($i=84.81^{+3.73}_{-5.27}$ degree), leading to a $M_2=1.12^{+0.10}_{-0.08} M_\odot$. For J064850, $i=50.82^{+13.52}_{-10.99}$ degree and 
$M_2=1.08^{+0.46}_{-0.26} M_\odot$.

\subsection{The nature of six candidates}

As mentioned in Section~\ref{Sec:4.2}, for J034813, J063350, and J064813, we constrain the orbital inclinations by modeling the ellipsoidal light curves using \texttt{PHOEBE}. The results show that the invisible objects' masses are around $1 M_{\odot}$. Therefore, we propose that the three binaries contain a massive WD, although the possibility of an NS cannot be ruled out. For J064717, J085102, and J104734, the orbital inclination cannot be constrained due to the absence of periodic light curves; therefore, we calculate the minimum mass of their invisible object with $i=90^\circ$. To constrain the inclination, one possible approach is to resolve the rotational broadening \citep{1994MNRAS.266..137M} of the tidally locked visible star using high-resolution spectra.

We cross-match the candidates with available X-ray surveys: \textit{Chandra} source catalog \citep{2010ApJS..189...37E}, \textit{XMM-Newton} source catalog \citep{2020A&A...641A.137T}, \textit{ROSAT} all-sky surveys \citep{1999A&A...349..389V} with a matching radius of 10\arcsec. We do not find reported X-ray observations for most of the sources, except for J085102. An X-ray source 2CXO J085102.0+114901 with an angular distance of 0.62\arcsec away was reported by \textit{Chandra}, which is a high probability X-ray counterpart of J085102. The X-ray flux is $F_{1.2-2.0\, \rm{keV}} = 6.67^{+3.53}_{-3.73} \times 10^{-16}\, \rm{erg}\, \rm{s}^{-1}\,\rm{cm}^{-2}$ in 1.2--2.0 keV band and $F_{0.5-7\, \rm{keV}} = 2.26^{+1.0}_{-1.0} \times 10^{-15}\, \rm{erg}\, \rm{s}^{-1}\,\rm{cm}^{-2}$ in 0.5--7.0 keV band. 
Using the distance of 0.78 kpc from \Gaia DR3 \citep{2022arXiv220800211G}, we convert the flux into the luminosity $L_{1.2-2.0\, \rm{keV}} = 4.87 \times 10^{28}\, \rm{erg}\, \rm{s}^{-1}$ and $
L_{0.5-7\, \rm{keV}} = 1.65 \times 10^{29}\, \rm{erg}\, \rm{s}^{-1}$, for 1.2--2.0 keV band and 0.5--7.0 keV band, respectively.
The X-ray luminosity of J085102 is $\sim 2-5$ orders of magnitude smaller than low-mass X-ray binaries in the quiescent state
\citep{2001NewAR..45..449L,2006ARA&A..44...49R,2014MNRAS.439.2771B}, which is about $10^{31-33} \rm{erg}\, \rm{s}^{-1}$, therefore J085102 is an X-ray faint source.
Notably, \citet{2020ApJ...902..114W} calculated the X-ray luminosity of J085102 in the 0.3--8 keV as $2 \times 10^{29}\, \rm{erg}\, \rm{s}^{-1}$, and the X-ray-to-bolometric luminosity ratio is ${\rm{log}}R_X = -4.84$, which was considered that the X-ray emission originates from normal stellar X-ray activity.

In addition to X-ray observations, we cross-match the six candidates with GALEX using a cross-matching radius of 20\arcsec. J063350 and J064717 are not observed by GALEX. 
J034813, J064850, J085102, and J104734 are covered in GALEX's footprints, 
but neither FUV nor NUV magnitudes are reported. 
The UV (non-)observations indicate the compact objects are intrinsically UV-faint; that is, they could be a cold WD or an NS. In the case of a WD, we use the GALEX detection limiting magnitude of NUV $\sim$ 25.5 mag \citep{2007ApJS..173..682M} and the WD cooling model \citep{2020ApJ...901...93B} to constrain the WD's NUV flux and obtain an upper limit of the effective temperature. The upper limits for J034813 and J064850 are 8000 K and 9500 K, respectively.

We select three binaries possibly containing unseen WDs by using the dynamical method, 
which is distinct from the spectroscopic decomposition method \citep{2014MNRAS.445.1331L,2016MNRAS.458.3808R} 
and the UV-excess selection method \citep{2016MNRAS.463.2125P}.
Since WDs in binaries are intrinsically faint relative to the bright MS stars (e.g. F-, G-, and K-type stars), the spectral decomposition method is typically limited to finding WD--M dwarf binaries \citep{2012MNRAS.419..806R,2017ApJ...850...34B}.
The UV-excess method can find WD--MS binaries with bright MS companions; 
however, stellar chromospheric UV emission \citep{2017ARA&A..55..159L} can severely contaminate the UV-excess candidates. Moreover, the UV-excess method cannot resolve UV-faint sources similar to our candidates J034813 and J064850.
In this aspect, the dynamical method is more robust since the orbital parameters and therefore the mass of the unseen compact object can be obtained.

\section{summary}\label{sec: summmary}

Based on the spectroscopic data from the LAMOST TD survey of four K2 plates \citep{2021RAA....21..292W}, we propose a sample consisting of six single-lined spectroscopic binary systems that may conceal compact objects. 
Multi-epoch (time-resolved) spectroscopic data provide solid measurements of the stellar radial velocities.
We utilize several methods to determine the orbital periods from the radial velocities, fit the radial velocity curve, and calculate the mass function. 
The six compact object candidates in our sample have $f(M_2) \gtrsim 0.1M_\odot$.

We use the SED fitting to estimate the stellar radius and mass. Based on the mass function and the mass of the visible star, we obtain the lower mass limit of the invisible component $M^{\rm{min}}_2$ assuming that $i=90^{\circ}$. 
The estimated $M^{\rm{min}}_2$ exceeds half of the visible stellar mass for all sources. 
We also obtain light curves from K2, ZTF, and \TESS surveys for three targets: J034813, J063350, and J064850, respectively.
These light curves show prominent ellipsoidal variability as expected from the tidal distortion of a compact object companion.
We use the \texttt{PHOEBE} model to constrain the orbital inclination angle and hence the mass of the unseen companion. The results indicate that these sources may be binary systems containing a massive white dwarf or neutron star.

In addition, we calculate the visible star's Roche lobe filling factor, showing that most of the systems are not filling up their Roche lobe and therefore could be non-accreting systems. 
The sources exhibit no X-ray detection except J085120, which is an X-ray faint source. Its X-ray emission can be attributed to stellar activity rather than accretion by the compact companion. 
We note that the six sources are worth follow-up observations to reveal their true nature. 

Thanks to various wide-field TD spectroscopic and photometric surveys, we expect the dynamical method to potentially unearth more hidden compact objects in binary systems.
Future observations of other K2 or \TESS footprints by LAMOST will aid in expanding the sample of compact objects with a wobbling stellar companion.

 \begin{acknowledgments}
 
We thank Jin-Bo Fu and Ling-Lin Zheng for their helpful discussions, and the anonymous referee for the constructive suggestions to improve the paper.
This work was supported by the National Key R\&D Program of China under grant 2021YFA1600401, the National Natural Science Foundation of China under grants 11925301, 11933004, 11988101, 12033006, 12103041, 12221003, and 12273057.
This paper uses the data from the LAMOST survey.
Guoshoujing Telescope (the Large Sky Area Multi-Object Fiber Spectroscopic Telescope LAMOST) is a National Major Scientific Project built by the Chinese Academy of Sciences. Funding for the project has been provided by the National Development and Reform Commission. LAMOST is operated and managed by the National Astronomical Observatories, Chinese Academy of Sciences.
This paper includes data collected by the \TESS mission and the K2 mission, which are obtained from the Mikulski Archive for Space Telescopes (MAST) at the Space Telescope Science Institute. The specific observations can be accessed via\dataset[10.17909/fwdt-2x66]{https://doi.org/10.17909/fwdt-2x66} \&\dataset[10.17909/T9WS3R]{https://doi.org/10.17909/T9WS3R}.
ZTF is a public-private partnership, with equal support from the ZTF Partnership and from the U.S. National Science Foundation through the Mid-Scale Innovations Program (MSIP).
 \end{acknowledgments}

 \software{
    astroARIADNE \citep{2022MNRAS.513.2719V},
    Astropy \citep{2013A&A...558A..33A,2018AJ....156..123A},
    Lightkurve \citep{2018ascl.soft12013L},
    Matplotlib \citep{Hunter:2007},
    NumPy \citep{harris2020array},
    Pandas \citep{mckinney-proc-scipy-2010},
    emcee \citep{2013PASP..125..306F},
    PHOEBE \citep{2005ApJ...628..426P,2011ascl.soft06002P,2020ApJS..250...34C}
  }

\bibliography{main}{}
\bibliographystyle{aasjournal}

\section*{Appendix}\label{sec:appendix}

To verify that the systems are single-lined spectroscopic binaries, we checked whether the CCF profile has a single peak, an indication of a single dominant component. We adopt two spectra for each target source to compute the CCF. These two spectra are chosen to be mostly in anti-orbital phases (near the quadrature phases), such that if there are two components, they would be easily detected in the velocity space. J104734 has poor SNR for most of the spectra, therefore we trade off the phase requirement for better SNR. J063350 and J064850 also have poor medium-resolution spectra, therefore we turn to use the low-resolution spectra. The results the Figure 5 suggest that these sources are most likely single-lined binaries given the current spectroscopic resolution.

Figure 6 presents the broadband SED fitting of three targets with ellipsoidal light curve variations. The fitting is performed using the \texttt{astroARIADNE} \citep{2022MNRAS.513.2719V} with a single stellar component model. The results suggest that these three targets can be well-fitted with the SED model; No evidence of contamination from other components supports our conclusions about the nature of these systems.

\begin{figure*}[h]
\centering
\includegraphics[scale=0.25]{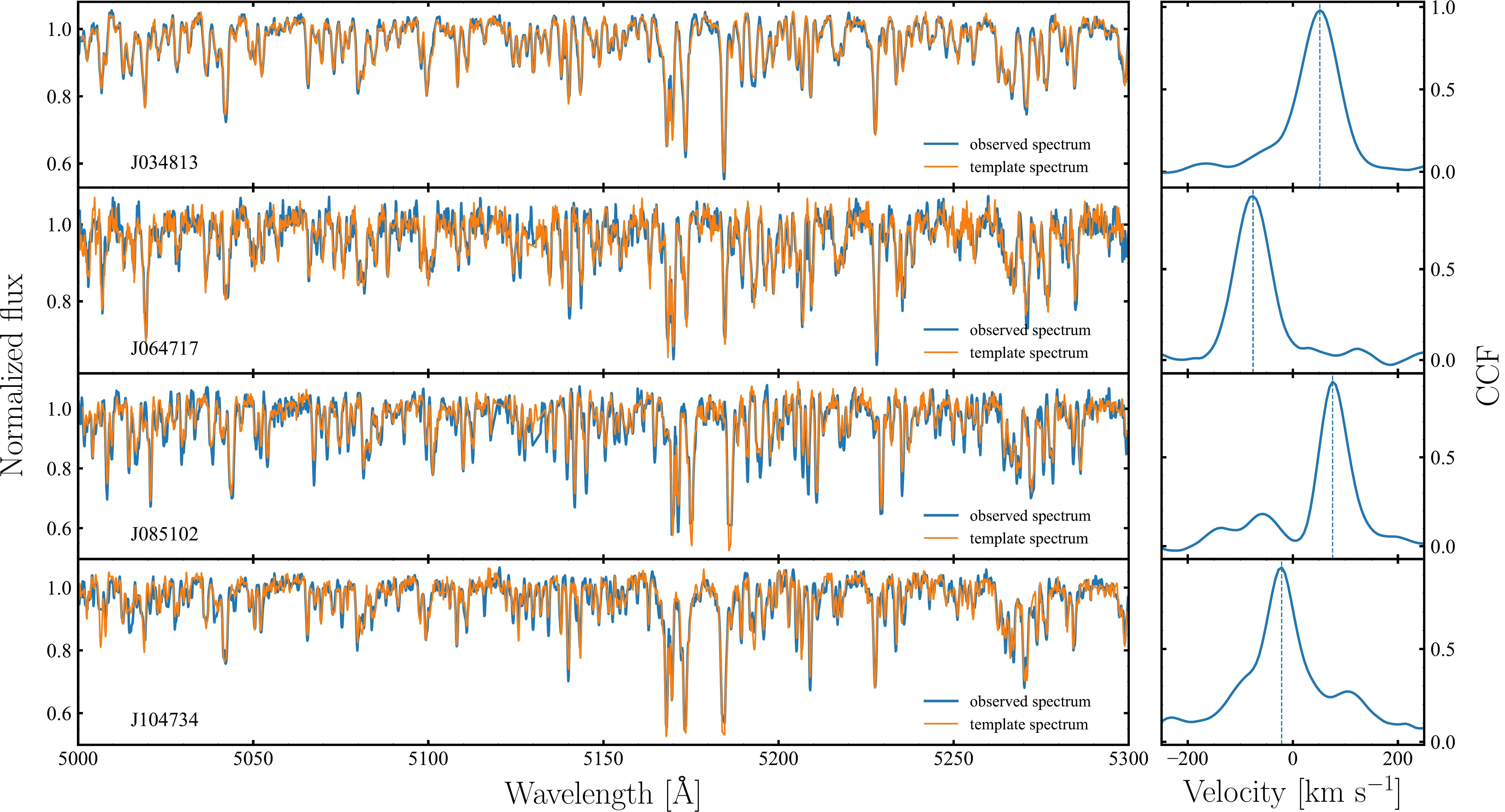}
\includegraphics[scale=0.25]{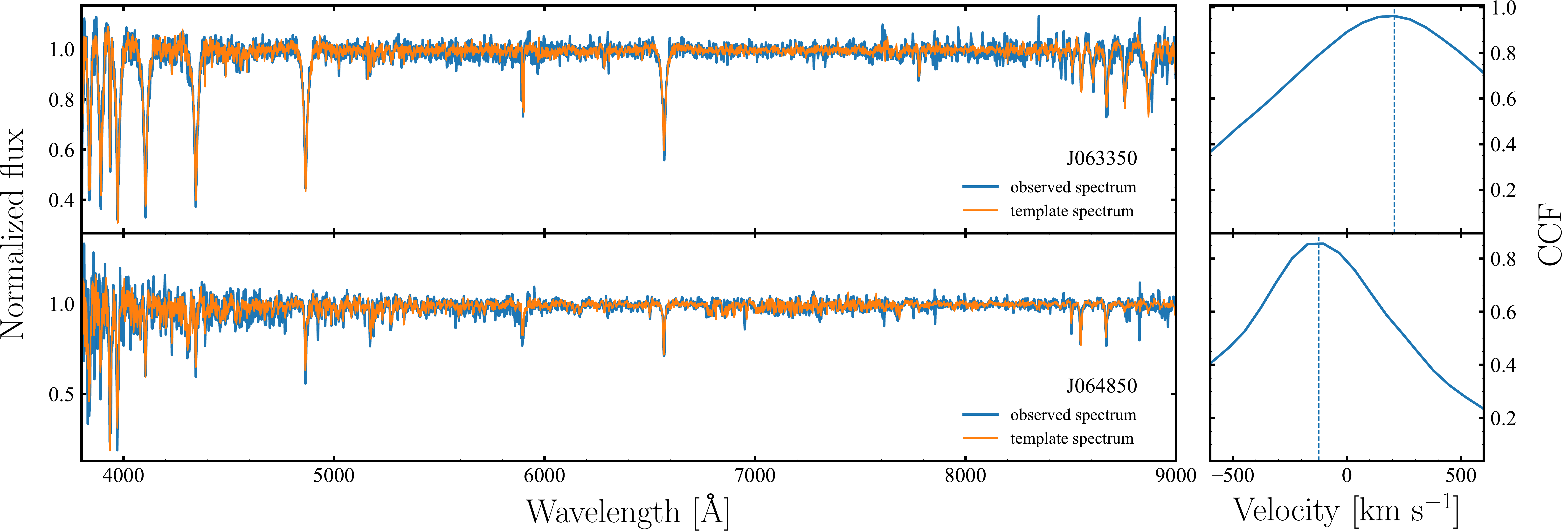}
\caption{LAMOST spectra (left panels) and CCF profile (right panels) for the candidates. In each left panel, two spectra used to perform the CCF are both observational spectra but are styled with distinct colors and labels to indicate which one is used as the template. The vertical dashed lines in the right panels indicate the measured (relative) radial velocity.}
\label{fig:CCF}
\end{figure*}

\begin{figure*}[ht]
\centering
\includegraphics[scale=0.35]{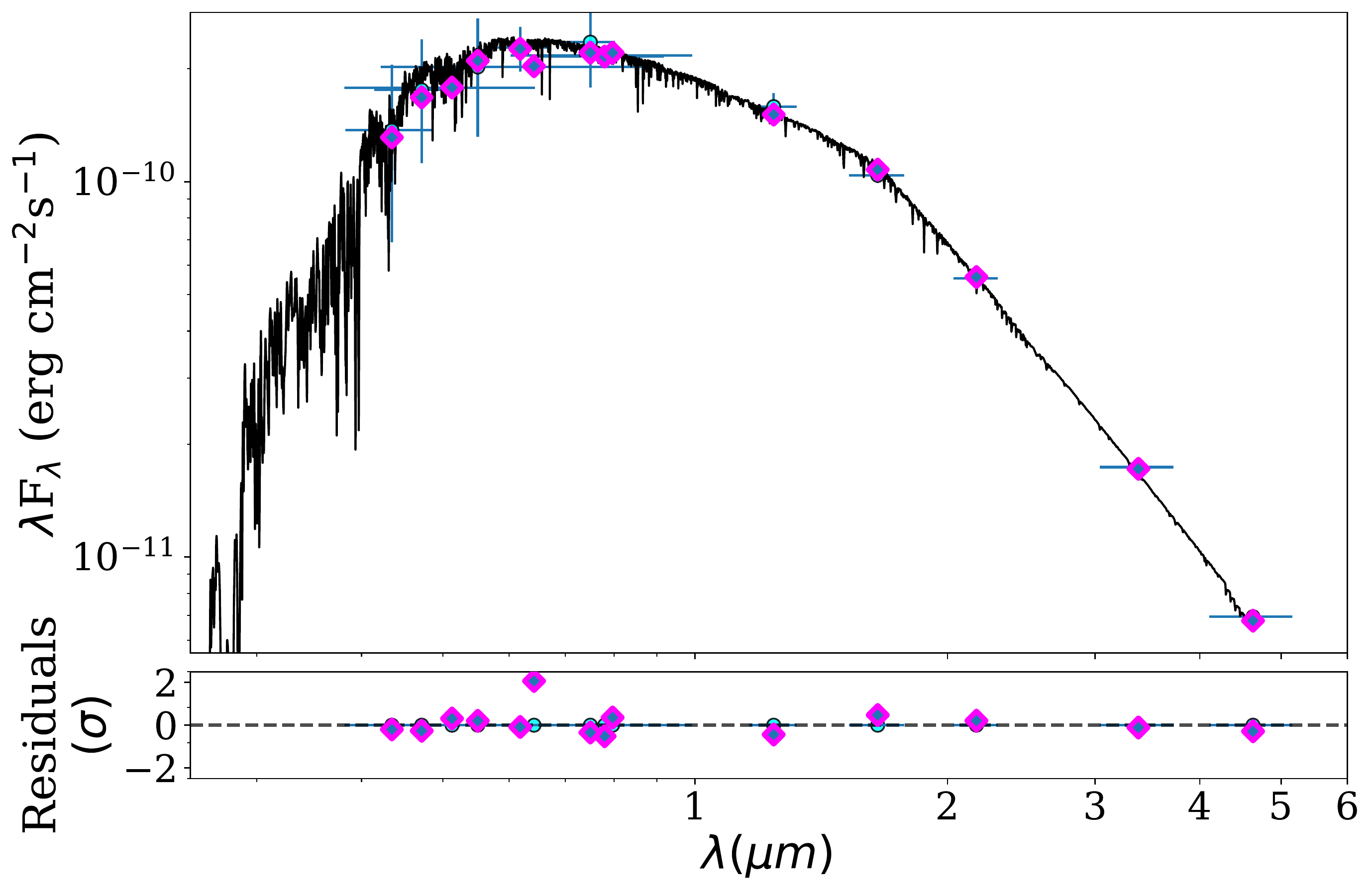}
\includegraphics[scale=0.35]{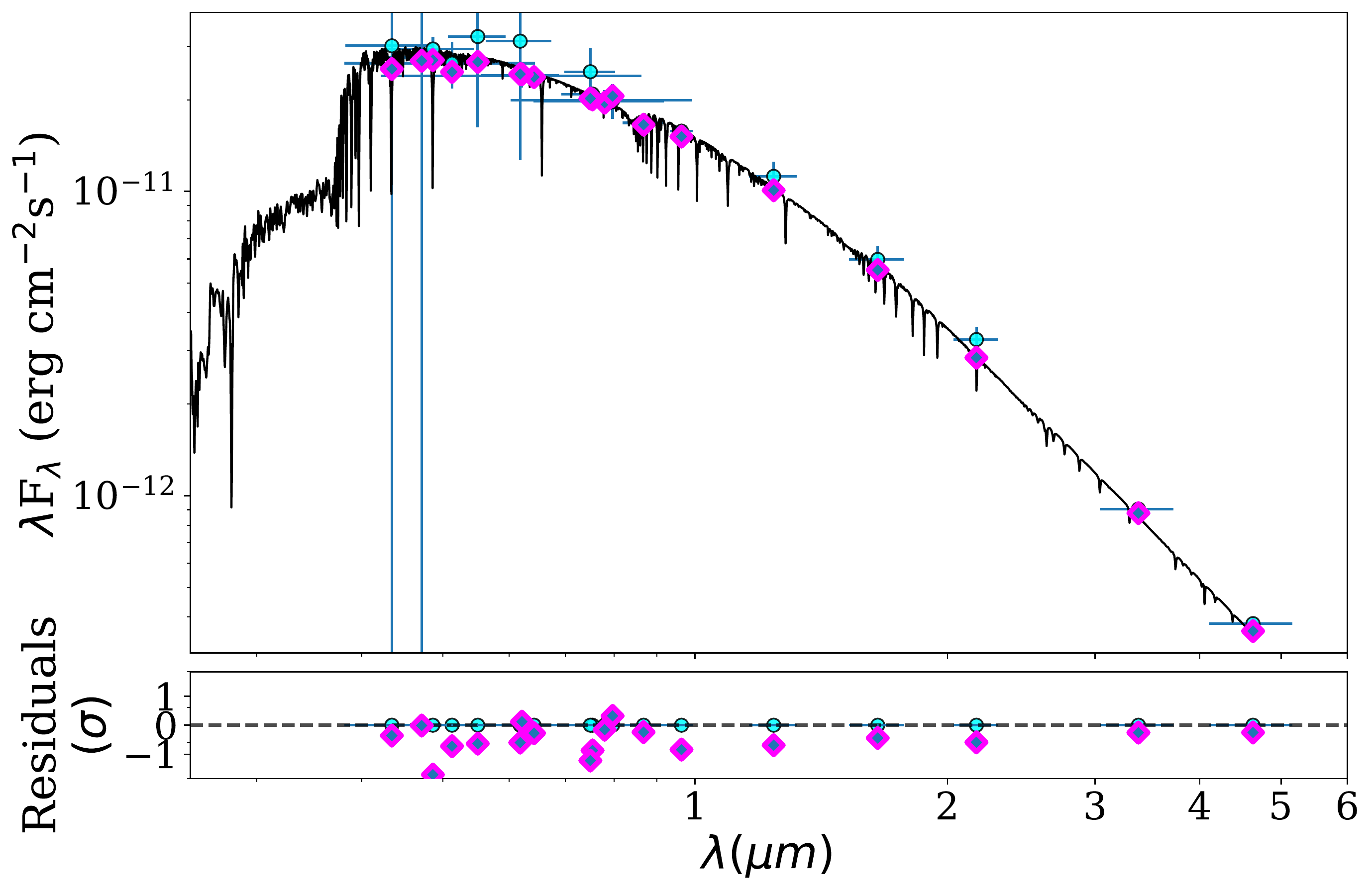}
\includegraphics[scale=0.35]{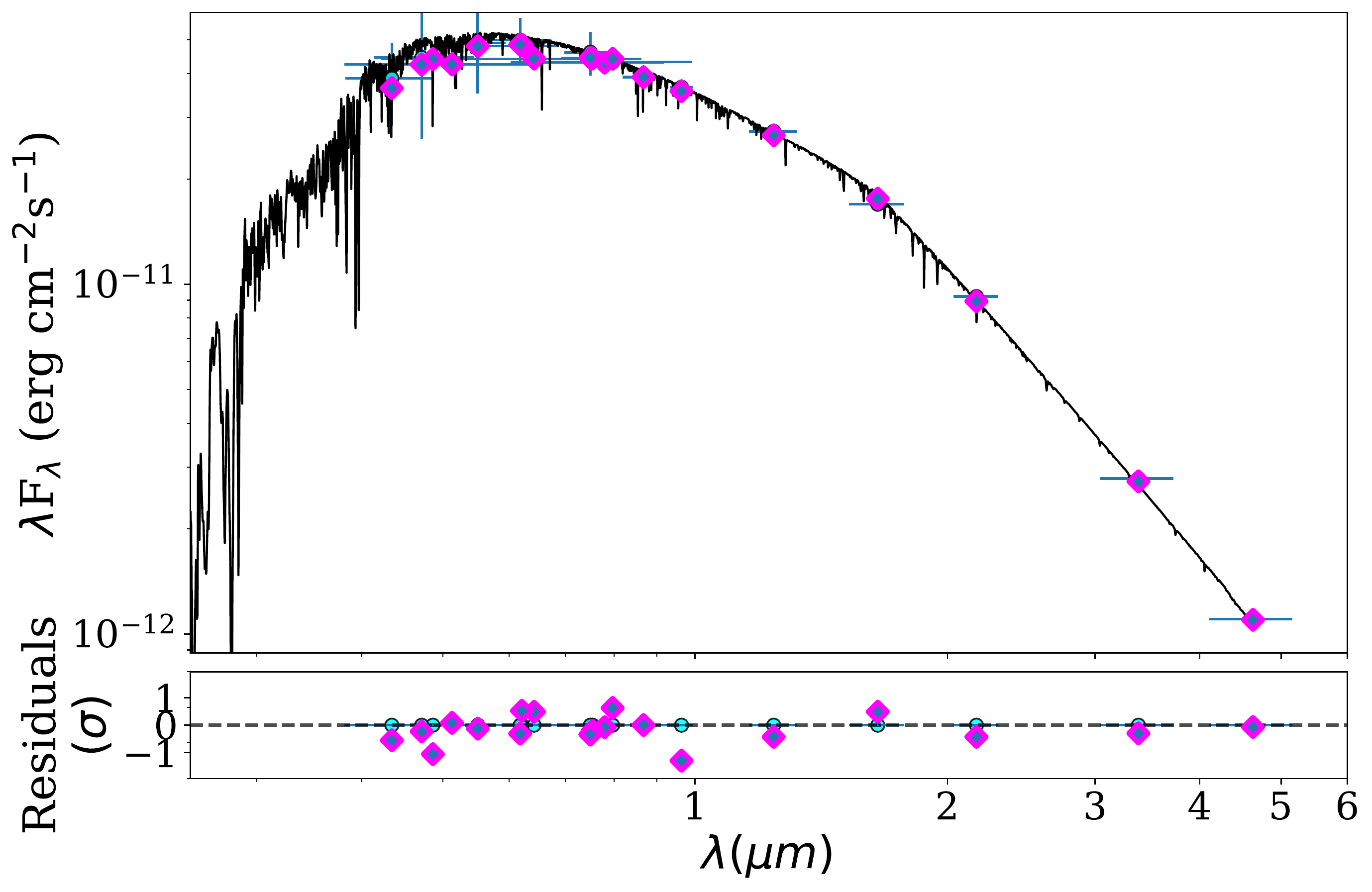}
\caption{The SED fitting results of J034813 (top panel), J063350 (middle panel), and J064850 (bottom panel). Cyan points are the observed fluxes, purple diamonds are the synthetic fluxes, and the black curve is the best-fit SED model.}
\label{fig:SED}
\end{figure*}

\end{document}